\documentclass[a4paper,aps,prl,twocolumn,floats,showpacs,superscriptaddress]{revtex4}
\usepackage{epsfig}

\newcommand{\avk}{\left< k \right>}
\newcommand{\fluck}{\left< k^2 \right>}
\newcommand{\condP}{P(k' \, \vert \, k)}
\newcommand{\condPN}[2]{P({#1} \, \vert \, {#2})}

\begin{document}

\title{Absence of epidemic threshold in scale-free networks 
with connectivity correlations}

\author{Mari{\'a}n Bogu{\~n}{\'a}} 
\affiliation{Departament de F{\'\i}sica Fonamental, Universitat de
  Barcelona, Av. Diagonal 647, 08028 Barcelona, Spain}
\author{Romualdo Pastor-Satorras}  
\affiliation{Departament  de F{\'\i}sica i Enginyeria Nuclear,
  Universitat Polit{\`e}cnica de Catalunya, Campus Nord,
  08034 Barcelona, Spain}
\author{Alessandro Vespignani}
\affiliation{The Abdus Salam International Centre for
Theoretical Physics, P.O. Box 586, Trieste,
I-34014, Italy}

\date{\today}

\begin{abstract}
  Random scale-free networks have the peculiar property of being prone
  to the spreading of infections.  Here we provide an exact result
  showing that a scale-free connectivity distribution with diverging
  second moment is a sufficient condition to have null epidemic
  threshold in unstructured networks with either assortative or
  dissortative mixing.  Connectivity correlations result therefore
  ininfluential for the epidemic spreading picture in these scale-free
  networks.  The present result is related to the divergence of the
  average nearest neighbors connectivity, enforced by the connectivity
  detailed balance condition.
\end{abstract}


\maketitle

Complex networks play a capital role in the modeling of many social,
natural, and technological systems which are characterized by peculiar
topological properties \cite{barabasi02,dorogorev}. In particular,
small-world properties \cite{watts98} and scale-free connectivity
distributions \cite{barab99} appear as common features of many
real-world networks. The statistical physics approach has been proved
a very valuable tool for the study of these networks, and several
surprising results concerning dynamical processes taking place on
complex networks have been repeatedly reported. In particular, the
absence of the percolation \cite{newman00,havlin01} and epidemic
\cite{pv01a,pv01b,lloydsir,virusreview} thresholds in scale-free (SF)
networks has hit the community because of its potential practical
implications.  The absence of the percolative threshold, indeed,
prompts to an exceptional tolerance to random damages
\cite{barabasi00}.  On the other hand, the lack of any epidemic
threshold makes SF networks the ideal media for the propagation of
infections, bugs, or unsolicited informations \cite{pv01a}.

Recent studies have focused in a more detailed topological
characterization of several social and technological networks. In
particular, it has been recognized that many of these networks
possess, along with SF properties, non-trivial connectivity
correlations \cite{assortative}. For instance, many social networks
show that nodes with high connectivity will connect more preferably to
highly connected nodes \cite{assortative}; a property referred to as
``assortative mixing''. On the opposite side, many technological and
biological networks show ``dissortative mixing''; \textit{i.e.} highly
connected nodes are preferably connected to nodes with low
connectivity \cite{alexei,alexei02,maslov02}.  Correlations are very
important in determining the physical properties of these networks
\cite{marian1} and several recent works are addressing the effect of
``dissortative mixing'' correlations in epidemic spreading
\cite{structured,sander,blanchard}. The fact that highly connected
nodes (hubs) are more likely to transmit the infection to poorly
connected nodes could somehow slow down the spreading process.  By
numerical simulations and analytical arguments it has been claimed
that, if strong enough, connectivity correlations might reintroduce an
epidemic threshold in SF networks, thus restoring the standard
tolerance to infections.

In this paper we analyze in detail the conditions for the lack of an
epidemic threshold in the susceptible-infected-susceptible model
\cite{anderson92} in SF networks. We find the exact result that {\em a
  SF connectivity distribution $P(k)\sim k^{-\gamma}$ with $2< \gamma\leq3$ in
  unstructured networks with assortative or dissortative mixing is a
  sufficient condition for a null epidemic threshold in the
  thermodynamic limit}.  In other words, the presence of two-point
connectivity correlations does not alter the extreme weakness of SF
networks to epidemic diffusion. This result is related to the
divergence of the nearest neighbors average connectivity, divergence
that is ensured by the connectivity detailed balance condition, to be
satisfied in physical networks. The present analysis can be easily
generalized to more sophisticated epidemic models.

In the following we shall consider unstructured undirected SF
networks, in which the only relevant property of the nodes is their
connectivity \cite{note_6}, with distribution $P(k)\sim C
k^{-\gamma}$, with $2< \gamma\leq3$.  $P(k)$ is defined as the probability that
a randomly selected node has $k$ connections to other nodes.  In this
case the network has unbounded connectivity fluctuations, signalled by
a diverging second moment $\fluck \to\infty$ in the thermodynamic limit
$k_c\to \infty$, where $k_c$ is the maximum connectivity of the network.
It is worth recalling that in growing networks $k_c$ is related to the
network size $N$ as $k_c\sim N^{1/(\gamma-1)}$ \cite{dorogorev}.  Finally,
we shall consider that the network presents assortative or
dissortative mixing allowing for non trivial two-point connectivity
correlations.  This corresponds to allow a general form for the
conditional probability, $\condP$, that a link emanated by a node of
connectivity $k$ points to a node of connectivity $k'$.

As a prototypical example for examining the properties of epidemic
dynamics in SF networks we consider the
susceptible-infected-susceptible (SIS) model \cite{anderson92}, in
which each node represents an individual of the population and the
links represent the physical interactions among which the infection
propagates.  Each individual can be either in a susceptible or
infected state.  Susceptible individuals become infected with
probability $\lambda$ if at least one of the neighbors is infected.
Infected nodes, on the other hand, recover and become susceptible
again with probability one. A different recovery probability can be
considered by a proper rescaling of $\lambda$ and the time. This model is
conceived for representing endemic infections which do not confer
permanent immunity, allowing individuals to go through the stochastic
cycle susceptible $\to$ infected $\to$ susceptible by contracting the
infection over and over again.  In regular homogeneous networks, in
which each node has more or less the same number of connections,
$k\simeq\avk$, it is possible to understand the behavior of the model by
looking at the average density of infected individuals $\rho(t)$ (the
prevalence).  It is found that for 
a spreading probability $\lambda\geq\lambda_c$,
where $\lambda_c$ is the epidemic threshold depending on the network
average connectivity and topology, the system reaches an endemic state
with a finite stationary density $\rho$.  
If $\lambda\leq\lambda_c$, the system
falls in a finite time in a healthy state with no infected individuals
($\rho=0$).

In SF networks the average connectivity is highly fluctuating and the
approximation $k\simeq\avk$ is totally inadequate. To take into account
the effect of the connectivity fluctuations, it has been shown that it
is appropriate to consider the quantity $\rho_k$
\cite{pv01a,pv01b,marian1}, defined as the density of infected nodes
within each connectivity class $k$. This description assumes that the
network is unstructured and that the classification of nodes according
only to their connectivity is meaningful \cite{note_6}.  Following
Ref.~\cite{marian1}, the mean-field rate equations describing the
system can be written as
\begin{equation}
  \frac{ d \rho_k(t)}{d t} =
  -\rho_k(t) +\lambda k \left[1-\rho_k(t) \right] \sum_{k'} 
    \condP \rho_{k'}(t).
  \label{generalized}
\end{equation}
The first term on the r.h.s. represents the annihilation of infected
individuals due to recovery with unitary rate. The creation term is
proportional to the density of susceptible individuals, $1-\rho_k$,
times the spreading rate, $\lambda$, the number of neighboring nodes, $k$,
and the probability that any neighboring node is infected. The latter
is the average over all connectivities of the probability $\condP
\rho_{k'}$ that a link emanated from a node with connectivity $k$ points
to an infected node with connectivity $k'$.  It is worth remarking
that, while keeping into account the two point connectivity
correlations, as given by the conditional probability $\condP$, yet we
have neglected higher order density-density and connectivity
correlations. Eq.~(\ref{generalized}) is therefore exact for the class
of Markovian networks \cite{marian1}, in the limit of low prevalence
($\rho(t) \ll 1$).

In the case of uncorrelated networks each link points, with
probability proportional to $k'P(k')$, to a node of connectivity $k'$,
regardless of the emanating node's connectivity.  In this case, in the
stationary state ($\partial_t\rho=0$), $\sum_{k'} \condP \rho_{k'}(t)$ assumes a
constant value independent on $k$ and $t$ and the system
(\ref{generalized}) can be solved self-consistently obtaining that the
epidemic threshold is given by \cite{virusreview}
\begin{equation}
  \lambda_c =\frac{\avk}{\fluck}.
  \label{eq:2}
\end{equation}
For infinite SF networks with $\gamma \leq 3$, we have $\fluck = \infty$, and
correspondingly $\lambda_c= 0$; \textit{i.e.} uncorrelated SF networks
allow a finite prevalence whatever the spreading rate $\lambda$ of the
infection.  Finally, from the solution of $\rho_k$, one can compute the
total prevalence $\rho$ using the relation $\rho=\sum_kP(k)\rho_k$.

In the case of correlated networks the explicit solution of
Eq.~(\ref{generalized}) is not generally accessible. However, it has
been shown that the epidemic threshold is given by \cite{marian1}
\begin{equation}
  \lambda_c = \frac{1}{ \Lambda_m},
  \label{eq:4}
\end{equation}
where $\Lambda_m$ is the largest eigenvalue of the \textit{connectivity
  matrix} $\mathbf{C}$, defined by $C_{k k'} = k\condP$.  In
Ref.~\cite{marian1} it has been shown how this general formalism
recovers previous results for uncorrelated networks, obtaining that,
in this case, $\Lambda_m=\fluck/\avk$. More generally, by looking at
Eq.~(\ref{eq:4}), the absence of an epidemic threshold corresponds to
a divergence of the largest eigenvalue of the connectivity matrix
$\mathbf{C}$ in the limit of an infinite network size $N\to\infty$.  In
order to provide some general statement on the conditions for such a
divergence we can make use of the Frobenius theorem for non-negative
irreducible matrices \cite{Gantmacher}. This theorem states the
existence of the largest eigenvalue of any non-negative irreducible
matrix, eigenvalue which is simple, positive, and has a positive
eigenvector. One of the consequences of the theorem is that it
provides a bound to such largest eigenvalue \cite{note_2}. In our case
the matrix of interest is the connectivity matrix and, since
$\mathbf{C}$ is non-negative and irreducible \cite{note_1}, it is
possible to find lower and upper bounds of $\Lambda_m$. In particular, we
can write \cite{note_2}
\begin{equation}
  \Lambda_m^2\geq \min_k \sum_{k'}\sum_{\ell}k'\ell \condPN{\ell }{k} \condPN{k'}{\ell}.
\label{frob}
\end{equation}
This inequality relates the lower bound of the largest eigenvalue
$\Lambda_m$ to the connectivity correlation function and, as we shall see,
allows to find a sufficient condition for the absence of the epidemic
threshold.

In order to provide an explicit bound to the largest eigenvalue we
must exploit the properties of the conditional probability $\condP$.
A key relation holding for all physical networks is that all links
must point from one node to another. This is translated in the
connectivity detailed balance condition \cite{marian1}
\begin{equation}
  k \condP P(k) = k' P(k \, \vert \, k') P(k'),
  \label{detbal}
\end{equation}
which states that the total number of links pointing from nodes with
connectivity $k$ to nodes of connectivity $k'$ must be equal to the
total number of links that point from nodes with connectivity $k'$ to
nodes of connectivity $k$.  This relation is extremely important since
it constraints the possible form of the conditional probability
$\condP$ once $P(k)$ is given.  By multiplying by a $k$ factor both
terms of Eq.~(\ref{detbal}) and summing over $k'$ and $k$, we obtain
\begin{equation}
 \fluck  = \sum_{k'}k'P(k')\sum_k kP(k \, \vert \, k'),
\label{detbal2}
\end{equation}
where we have used the normalization conditions $\sum_k P(k) = \sum_{k'}
\condP = 1$.  The term $\overline{k}_{nn}(k', k_c)= \sum_k kP(k \, \vert
\, k')$ defines the average nearest neighbor connectivity (ANNC) of
nodes of connectivity $k'$.  This is a quantity customarily measured
in SF and complex networks in order to quantify degree-degree
correlations \cite{alexei,alexei02,maslov02}.  The dependence on $k_c$
is originated by the upper cut-off of the $k$-sum and it must be taken
into account since it is a possible source of divergences in the
thermodynamic limit. In SF networks with $2< \gamma < 3$ we have that the
second moment of the connectivity distribution diverges as $\fluck\sim
k_c^{3-\gamma}$ \cite{note_3}.  We thus obtain that
\begin{equation}
  \sum_{k'}k'P(k') \overline{k}_{nn}(k',k_c) 
  \simeq \frac{C}{(3-\gamma)} k_c^{3-\gamma}.
\label{detbal3}
\end{equation}
In the case of dissortative mixing \cite{assortative}, the function
$\overline{k}_{nn}(k',k_c)$ is decreasing with $k'$ and, since
$k'P(k')$ is an integrable function, the l.h.s. of Eq.~(\ref{detbal3})
has no divergence related to the sum over $k'$. This implies that the
divergence must be contained in the $k_c$ dependence of
$\overline{k}_{nn}(k',k_c)$.  In other words, the function
$\overline{k}_{nn}(k',k_c)\to\infty$ for $k_c\to\infty$ in a non-zero measure
set. In the case of assortative mixing, $\overline{k}_{nn}(k',k_c)$ is
an increasing function of $k'$ and, depending on its rate of growth,
there may be singularities associated to the sum over $k'$. Therefore,
this case has to be analyzed in detail. Let us assume that the ANNC
grows as $\overline{k}_{nn}(k',k_c) \simeq \alpha k'^{\beta}$, $\beta>0$, when $k'
\to \infty$. If $\beta <\gamma -2$, again there is no singularity related to the
sum over $k'$ and the previous argument for dissortative mixing holds.
When $\gamma -2 \leq \beta <1$ there is a singularity coming from the sum over
$k'$ of the type $\alpha k_c^{\beta-(\gamma-2)}$. However, since
Eq.~(\ref{detbal3}) comes from an identity, the singularity on the
l.h.s. must match both the exponent of $k_c$ and the prefactor on the
r.h.s.  In the case $\gamma -2 \leq \beta <1$, the singularity coming from the
sum is not strong enough to match the r.h.s. of Eq.~(\ref{detbal3})
since $\beta-(\gamma-2) < 3-\gamma$. Thus, the function
$\overline{k}_{nn}(k',k_c)$ must also diverge when $k_c \to \infty$ in a
non-zero measure set. Finally, when $\beta >1$ the singularity associated
to the sum is too strong, forcing the prefactor to scale as $\alpha \simeq r
k_c^{1-\beta}$ and the ANNC as $\overline{k}_{nn}(k',k_c)\simeq r k_c^{1-\beta}
k'^{\beta}$. It is easy to realize that $r \leq 1$, since the ANNC cannot
be larger than $k_c$. Plugging the $\overline{k}_{nn}(k',k_c)$
dependence into Eq.~(\ref{detbal3}) and simplifying common factors, we
obtain the identity at the level of prefactors
\begin{equation}
  \frac{r}{2-\gamma+\beta} = \frac{1}{3-\gamma}.
  \label{eq:pref}
\end{equation}
Since $\beta>1$ and $r<1$, the prefactor in the l.h.s. of
Eq.~(\ref{eq:pref}) is smaller than the one of the r.h.s. This fact
implies that the tail of the distribution in the l.h.s. of
Eq.~(\ref{detbal3}) cannot account for the whole divergence of its
r.h.s. This means that the sum is not the only source of divergences
and, therefore, the ANNC must diverge at some other point \cite{note_5}.

The large $k_c$ behavior of the ANNC can be plugged in
Eq.~(\ref{frob}) obtaining that
\begin{equation}
\Lambda_m^2\geq \min_k\sum_{\ell} \ell \condPN{\ell}{k} \overline{k}_{nn}(\ell ,k_c)
\label{frob2}
\end{equation}
The r.h.s. of this equation is a sum of positive terms and diverges
with $k_c$ at least as $\overline{k}_{nn}(\ell ,k_c)$ both in the
dissortative or assortative cases \cite{note_4}.  This readily implies
that $\Lambda_m\geq\infty$ for all networks with diverging $\fluck$.  Finally
Eq.~(\ref{eq:4}) yields that the epidemic threshold vanishes in the
thermodynamic limit in all SF networks with assortative and
dissortative mixing if the connectivity distribution has a diverging
second moment; {\em i.e. a SF connectivity distribution with exponent
  $2<\gamma\leq3$ is a sufficient condition for the absence of an epidemic
  threshold in unstructured networks with arbitrary two-point
  connectivity correlation function}.

In physical terms, the absence of the epidemic threshold is related to
the divergence of the average nearest neighbors connectivity $\left<
  \overline{k}_{nn}\right>_{N}$ in SF networks. This function is
defined by
\begin{equation}
  \left< \overline{k}_{nn}\right>_{N}=\sum_k P(k) \overline{k}_{nn}(k ,k_c),
\end{equation} 
where we have explicitly considered $k_c$ as a growing function of the
network size $N$.  By using the analysis shown previously it follows
that $\left< \overline{k}_{nn}\right>_{N}\to\infty$ when $N\to\infty$.  In SF
networks this parameter takes into account the level of connectivity
fluctuations and appears as ruling the epidemic spreading dynamics.
Somehow the number of neighbors that can be infected in successive
steps is the relevant quantity. Only in homogeneous networks, where
$\left< \overline{k}_{nn}\right>_{N}\simeq\avk$, the epidemic spreading
properties can be related to the average connectivity.  Noticeably,
the power-law behavior of SF networks imposes a divergence of $\left<
  \overline{k}_{nn}\right>_{N}$ independently of the level of
correlations present in the network.  This amounts to lower to zero
the epidemic threshold.  On the practical side, connectivity
correlation functions can be measured in several networks and show
assortative or dissortative behavior depending on the system. These
measurements are always performed in the presence of a finite $k_c$
that allows the regularization of the function $\overline{k}_{nn}(k
,k_c)$.  The most convenient way to exploit the infinite size
singularity is to measure the average nearest neighbor connectivity
for increasing network sizes.  All SF networks with $2<\gamma\leq3$ must
present a diverging $\left< \overline{k}_{nn}\right>_{N}$ for $N\to\infty$.
This statement is independent of the structure of the correlations
present in the networks.

It is worth stressing that the divergence of $\left<
  \overline{k}_{nn}\right>_{N}$ is ensured by the connectivity
detailed balance condition alone. Thus it is a very general results
holding for all SF networks with $2<\gamma\leq3$.  On the contrary, the SF
behavior with $2<\gamma\leq3$ is a necessary condition for the lack of
epidemic threshold only in networks with general two-point
connectivity correlations and in absence of higher-order correlations.
The reason is that the relation between the epidemic threshold and the
maximum eigenvalue of the connectivity matrix only holds for these
classes of networks. Higher order correlations, or the presence of an
underlying metric in the network \cite{note_6}, can modify the rate
equation at the basis of the SIS model and may invalidate the present
discussion.

\begin{acknowledgments}
  This work has been partially supported by the European commission
  FET Open project COSIN IST-2001-33555. R.P.-S. acknowledges
  financial support from the Ministerio de Ciencia y Tecnolog{\'\i}a
  (Spain).
\end{acknowledgments}


\begin{thebibliography}{27}
\bibitem[{not({\natexlab{b}})}]{note_2}
\bibinfo{note}{From the Frobenius theorem \cite{Gantmacher} it can be proved
  that the maximum eigenvalue, $\Lambda_m$, of any non-negative irreducible
  matrix, $A_{k k'}$, satisfies the inequality $ \Lambda_m \geq \min_k
  \frac{1}{\psi(k)}\sum_{k'} A_{k k'} \psi(k')$, where $\psi(k)$ is any
  positive vector. By setting ${\mathbf A}={\mathbf C^2}$ and $\psi(k)=k$ we
  recover the inequality of Eq.~(\ref{frob}).}
\bibitem[{not({\natexlab{c}})}]{note_1}
\bibinfo{note}{The irreducible property of the connectivity matrix is a simple
  consequence of the fact that all the connectivity classes in the network are
  accessible. That is, starting from the connectivity class $k$ it is always
  possible to find a path of links that connects this class to any other class
  $k'$ of the network. If this is not the case it means that the network is
  built up of disconnected irreducible subnetworks and, therefore, we can apply
  the same line of reasoning to each subnetwork. Notice that being irreducible
  is not equivalent to being fully connected at the node to node level, but at
  the class to class level.}
\bibitem[{not({\natexlab{d}})}]{note_3}
\bibinfo{note}{For $\gamma=3$ the second moment diverges as $\fluck \sim \ln
  k_c$ but the argument, though more involved, is still valid.}
\bibitem[{not({\natexlab{e}})}]{note_5}
\bibinfo{note}{The case $\beta=1$, that is, $ \overline{k}_{nn}(k' ,k_c) \sim
  \alpha k'$ when $k' \to \infty$, is rather pathological because is the only
  case where the ANNC can be a non diverging function. If $\alpha<1$,
  Eq.~(\ref{detbal3}) implies that it must exist at least one diverging point
  out of the one coming from the sum over $k'$ and the discussion made in the
  text is valid. However, when $\alpha=1$ this divergence is enough to fulfill
  the identity (\ref{detbal3}) and $\overline{k}_{nn}(k ,k_c)$ can be a
  convergent function when $k_c \to \infty$. Even in this pathological
  situation it is possible to prove that the maximum eigenvalue diverges in the
  thermodynamic limit. In order to do so the argumentation is to be translated
  to the second moment of the nearest neighbor's connectivity. Since the
  inequality $\langle x^2 \rangle \geq \langle x \rangle^2$ is true for any
  random variable, the function $\overline{k^2}_{nn}(k ,k_c)$ is either $\sim
  k'^2$, in which case the maximum eigenvalue is trivially $\infty$, or the
  probability $P(k' \, | \, k)$ approaches to SF distribution when $k_c \to
  \infty$ and $\Lambda_m$ diverges as well.}
\bibitem[{not({\natexlab{f}})}]{note_4}
\bibinfo{note}{One may argue that, since we are calculating a minimum for $k$,
  if the transition probability $P(\ell \, | \, k_0)$ is zero at some point
  $k_0$, this minimum is zero. In this case it is possible to show that
  repeating the same argument with ${\mathbf C}^3$ instead of ${\mathbf C}^2$
  provides us an inequality that avoids this problem.}
\end{thebibliography}
\end{document}